\documentclass[11pt,fleqn,twoside]{article}
\usepackage{latexsym}

\newcommand{\name}[1]{\begin{flushleft}
                       \LARGE \bf #1
                       \end{flushleft}}

\newcommand{\Author}[1]{\begin{flushleft}
                       \it #1 \end{flushleft}}

\newcommand{\Adress}[1]{\begin{flushleft}
                       \it #1 \end{flushleft}}

\newcommand{\ehkol}{Author \ name}
\newcommand{\ohkol}{Article \ name}

\newcommand{\be}{\begin{equation}}
\newcommand{\ee}{\end{equation}}
\newcommand{\ba}{\hspace*{-5pt}\begin{array}}
\newcommand{\ea}{\end{array}}

\newcommand{\ds}{\displaystyle}
\makeatletter
\renewcommand{\@evenhead}
{\hspace*{-3pt}\raisebox{-15pt}[\headheight][0pt]{\vbox{\hbox to
\textwidth {\thepage \hfil \ehkol}\vskip4pt \hrule}}}
\renewcommand{\@oddhead}{
\hspace*{-3pt}\raisebox{-15pt}[\headheight][0pt]{\vbox{\hbox to
\textwidth {\ohkol \hfil \thepage}\vskip4pt\hrule}}}
\renewcommand{\@evenfoot}{}
\renewcommand{\@oddfoot}{}
\makeatother

\begin{document}

\setcounter{page}{331}
\renewcommand{\ehkol}{A. Sergeyev}
\renewcommand{\ohkol}{On Parasupersymmetries in a Relativistic Coulomb
Problem for}
\thispagestyle{empty}
\begin{flushleft}
\footnotesize \sf
Symmetry in Nonlinear Mathematical Physics \qquad 1997, V.2,
\pageref{sergeyev-fp}--\pageref{sergeyev-lp}.
\end{flushleft}

\name{On Parasupersymmetries in \\ Relativistic Coulomb Problem
for
\\ the Modif\/ied Stueckelberg Equation}\label{sergeyev-fp}

\Author{Arthur SERGEYEV\footnote{also known as Artur SERGYEYEV}}

\Adress{Institute of Mathematics of the National Academy of
Sciences of Ukraine, \\ 3 Tereshchenkivs'ka Str.,  Kyiv 4,
Ukraine\footnote{Current (2002) address: Silesian University in
Opava, Mathematical Institute, Bezru\v covo n\'am. 13, 746~01
Opava, Czech Republic\\ E-mail: {\tt arthurser@imath.kiev.ua,
Artur.Sergyeyev@math.slu.cz}}}

\begin{abstract}
\noindent We consider the Coulomb problem for the modif\/ied
Stueckelberg equation. For some specif\/ic values of parameters,
we establish the presence of parasupersymmetry for spin-1 states
in this problem and give the explicit form of corresponding
parasupercharges.
\end{abstract}

\section*{Introduction}
In spite of the striking progress of quantum f\/ield theory (QFT) during
last three decades, the problem of description of bound states in QFT context
isn't yet completely solved. This state of things stimulates the use
of various approximate methods in order to obtain physically
important results.

In this paper, we  work within the frame of the so-called
one-particle approximation (OPA) which consists in neglecting
creation and annihilation of particles
and the quantum nature of external f\/ields which the only particle we
consider interacts with. This approximation is valid for the case
where the energy of the interaction between the particle and the
f\/ield is much less than the mass of the particle. In OPA, the
operator of the quantized f\/ield corresponding to our particle
is replaced by the "classical" (i.e., non-secondly quantized) quantity
which is in essence nothing but the matrix
element of that operator between vacuum and one-particle state
(cf.~[1] for the case of spin $1/2$ particles). This quantity
may be interpreted as a wave function of the particle
and satisf\/ies a {\it linear } equation, in
which external f\/ield also appears as a classical quantity.

But for the case of massive charged spin-1 particles interacting
with external electromagnetic f\/ield, even such relatively simple
equations lead to  numerous dif\/f\/iculties and inconsistencies
[2, 3]. Only in 1995 Beckers, Debergh and Nikitin [2] overcame
most of them and suggested a new equation describing spin-1
particles. They exactly solved it and pointed out its
parasupersymmetric properties for the case of external constant
homogeneous magnetic f\/ield [2]. (To f\/ind out more about
parasupersymmetry, refer to [4] and references therein.)

The evident next step is to study this model for another
physically interesting case of  external Coulomb f\/ield (and in
particular to check the possibility of existence of the
parasupersymmetry in this case too).

In order to overcome some dif\/f\/iculties arising in the process
of such a study, we consider in [3] and here a "toy" model
corresponding to a particle with two possible spin states: spin-1
and spin-0. Beckers, Debergh and Nikitin in [2] suggested a new
equation for this case too, but our one is slightly more general
(refer to Section 1).

The plan of the paper is as follows. In Section 1, we write down the
equation
describing our model (we call it modif\/ied Stueckelberg equation).
In Section 2,
we brief\/ly recall the results from [3] concerning the exact
solution of this
model for the case of the Coulomb f\/ield of attraction. Finally, in
Section 3, we
point out the existence of the parasupersymmetry in
the Coulomb f\/ield for spin-1
states at  particular values of some parameters of our model.

\section{Modif\/ied Stueckelberg equation}
We consider the so-called modif\/ied Stueckelberg
equation written in
the second-order for\-ma\-lism [3]:
\begin{equation}
(D_{\mu}D^{\mu}+m_{\rm eff}^{2}) B^{\nu}+iegF_{  \rho}^{\mu}
B^{\rho} = 0,
\end{equation}
where
\begin{equation}
m_{\rm eff}^{2} =M^{2} +k_{2} \mid e^{2} F_{\mu \nu} F^{\mu \nu}
\mid ^{1/2}, \quad D_ \mu = \partial _\mu + ieA_\mu, \quad
 F_{\mu \nu} =\partial _\mu
A_\nu -\partial _\nu A_\mu .
\end{equation}

 We use here the $\hbar =c=1$
units system and following notations: small Greek letters denote
indices
which refer to the Minkowski 4-space and run from 0 to 3 (unless
otherwise stated);
we use the following metrics of the Minkowski space: $g_{\mu \nu}
={\rm diag}[1,-1,-1,-1]$; the four-vector is written as
$N^{\mu} =(N^{0},{\bf
N})$, where the bold letter denotes its three-vector part;
coordinates and
derivatives are $x^{\mu}=(t,{\bf r})$, $ \partial _\mu
= \partial / \partial x^{\mu}$; $A^{\mu}$  are potentials of external
electromagnetic f\/ield;
$e,g$ and $M$ are, respectively, charge, gyromagnetic ratio
and mass of
the particle described by (1). Its wave function is given by the
four-vector $B^{\mu}$. In the free $(e=0)$ case [2] this particle has
two possible
spin states: spin-0 and spin-1 ones with the same mass $M$.

Equation (1) generalizes the modif\/ication of the Stueckelberg
equation from [2] for the case of an arbitrary gyromagnetic ratio
$g$ (authors of [2] consider only the $g=2$ case). We also put the
module sign at  expression (2) for $m_{\rm eff} ^{2}$ (in spite of
[2]) in order to avoid complex energy eigenvalues in the Coulomb
problem (otherwise $\mu_{i}$ (9) and, hence, energy eigenvalues of
the discrete spectrum $E^{inj}$ will be complex).

\section{Coulomb f\/ield}
The 4-potential, corresponding to
the Coulomb f\/ield of
attraction, is:
\begin{equation}
{\bf A} = 0, \quad A^{0}= - Z e /r, \quad Z > 0.
\end{equation}
 Since it is static and spherically symmetric, energy $E$ and total
momentum ${\bf J}={\bf L}+{\bf S}$  (${\bf L}$~is an angular
momentum and
${\bf S}$ is a spin) are integrals of motion.
Let us decompose the wavefunctions of stationary states with
f\/ixed energy
$E$ in the basis of
common eigenfunctions of ${\bf J}^{2}$ and ${\bf J}_{z}$
with eigenvalues $j(j+1)$ and $m$, respectively, $j=0,1,2,\ldots$, for
a given $j$, \ $m = -j, -j+1, \ldots ,j$. The corresponding
eigenmodes are:
\begin{equation}
\ba{l}
\ds  B^{0} _{Ejm}=i F_{Ej} (r)Y_{jm}\exp(-iEt) ,\\[3mm]
\ds {\bf B}_{Ejm} =
\exp(-iEt) \sum\limits_{\sigma = -1,0,1} B^{(\sigma)} _{Ej}(r)
{\bf Y}^{(\sigma)}_{jm},
\ea
\end{equation}
where ${\bf Y}^{(\sigma)}_{jm}$ are spherical vectors (see [5] for
their explicit form) and $Y_{jm}$ are usual spherical functions.
It may be shown that $F_{Ej}$ and $B^{\sigma} _{Ej}$, $(\sigma =
-1,0,1)$ in fact don't depend on $m$ [5]. For the sake of brevity
from now on we shall often suppress the indices $Ej$ at $F_{Ej}$
and $B^{(\sigma)} _{Ej}\  (\sigma = -1,0,1)$.

The substitution of (3) and (4) into (1) yields the following
equations for $F$ and $B^{(\sigma)}$:
\begin{equation}
 {\rm T} V =(2/r^{2}){\rm P} V ,\qquad {\rm T} B^{(0)} = 0,
\end{equation}
where
\[ V =(F  B^{(-1)}  B^{(1)})^{\dagger},\qquad
{\rm P}=\left ( \begin{array}{ccc} 0 & -b & 0 \\ b & 1 & -a \\0 & -a
& 0\end{array} \right)
\]
($ ^{\dagger}$ denotes matrix transposition, $a=
\sqrt{j(j+1)}$,\quad $\beta = Z e^{2}$, $b=\beta g / 2 $);
\begin{equation}
{\rm T}=(E+ \beta / r)^{2} +d^{2}/d r^{2}
+ (2/r) d/d r -j(j+1)/ r^{2}
- M^{2} - k_2 \beta / r^{2}.
\end{equation}
 Notice that for $j=0$\  $B^{(0)}_{Ej}=B^{(1)} _{Ej} \equiv 0$ [5].

By the appropriate replacement of the basis (introducing new
unknown functions $K^{(i)}$, $i=1,2,3$, being linear combinations
of $B^{(\sigma )}$, $\sigma = -1,1$ and $F(r)$), equations (5) for
radial functions may be reduced to the following form [3] (for
convenience we denote $K^{(0)} \equiv B^{(0)}$):
\begin{equation}
{\rm T} K^{(i)}=(2 \lambda _{i} / r^{2}) K^{(i)} \qquad i=0,1,2,3,
\end{equation}
  where
\begin{equation}
\lambda_{l}=1/2 + (-1)^{l-1} \sqrt{(j+1/2)^{2} - (\beta g / 2)^{2}},
\qquad l=1,2,\quad
 \lambda_{0} = \lambda_{3} = 0.
\end{equation}

Energy eigenvalues of the discrete spectrum for (7) are [3]:
\begin{equation}
 E^{inj} =M/ \sqrt{1+\beta ^{2}/(n+ \mu _{i}+1)^{2}},
\end{equation}
where  $n=0,1,2, \dots$; $j=0,1,2 \dots$; $i=0,1,2,3$ and
\begin{equation}
\mu_{i} =-1/2+\sqrt{(j+1/2)^{2} - \beta ^{2} + 2 \lambda _{i} + k_{2}
\beta }
\end{equation}
(index $i$ corresponds to the following eigenmode: $K^{(i)} \neq 0$,
other
functions $K^{(l)}=0$; since for $j=0$ $K^{(0)} = K^{(3)} \equiv 0$,
in this case, we have only two branches corresponding to $i=1,2$).
 Branches of the spectrum for $i=0$ and $i=3$ are completely
identical, i.e., we meet here a twofold degeneracy.

The discrete spectrum eigenfunctions are [3]
\begin{equation}
 K^{(i)nj} = c^{inj}x^{\mu_{i}}\exp(-x/2) L_{n}^{\mu _{i}}(x),
\end{equation}
where $c^{inj}$ are normalization constants, $x=2r \sqrt{
M^{2}-E^{2}}$, $L^{\alpha} _{n}$ are Laguerre polynomials,
 $n=0,1,2, \dots$; $j=0,1,2 \dots$; $i=0,1,2,3 $.

\section{Parasupersymmetry}

We consider the case $k_2=0, g=2, j > 0$ where [3]
\[
\mu _{1}=\lambda _{1}, \quad \mu _{0}=\lambda _{1} - 1,\quad
\mu_{2}=\lambda_{1} - 2
\]
and hence the above energy eigenvalues (9) possess
the threefold extra degeneracy:
\begin{equation}
 E^{1,n+1,j} = E^{0,n,j} = E^{3,n,j} = E^{2,n-1,j}, \quad n>1.
\end{equation}
Moreover, we restrict ourselves by considering only spin-1 states,
setting the condition (compatible with (1) for the case of Coulomb
f\/ield if $k_2=0$, $g=2$ for the states with $j > 0$ [3])
\begin{equation}
\ba{l}
\ds D_\mu B^{\mu} _{Ejm} = \Bigl\{ E K^{(3)} _{Ej} +
\sum\limits_{i=1} ^{2} \Bigl[ d K
^{(i)} _{Ej} /dr + \\[4mm]
\ds \qquad \qquad  + (1 + \lambda _i) K^{(i)} _{Ej} /r - E \beta
K^{(i)} _{Ej}/ \lambda _i \Bigr]  \Bigr\}\exp (-i E t) Y_{jm} = 0.
\ea
\end{equation}
We use here the term "spin-1 states" by the analogy with the free
$(e=0)$ case (cf.~[1]), i.e., we call spin-1 states the states
satisfying (13).
 In virtue of (13) the component $K^{(3)}$ is expressed via
 the remaining
ones and isn't independent anymore [3]. This implies that we must
deal with only three branches of the discrete spectrum $E^{inj}$,
$i=0,1,2$.

Equations (7) which remain after the exclusion of $K^{(3)}$ may
be rewritten in the following form:
\begin{equation}
{\rm H }\psi = \varepsilon \psi,
\end{equation}
where
\begin{equation}
\ba{l}
\psi = (K^{(1)} K^{(0)} K^{(2)})^{ \dagger},\quad
{\rm H}=2 {\rm diag\; \lbrack H_1, H_0, H_2 \rbrack }; \\[2mm]
{\rm H}_i =- d^{2}/dr^{2} - (2/r) d/dr
- \lambda _i (\lambda _i + 1)/
r^{2} - 2 \beta
E /r + \\[2mm]
\qquad \qquad + (1/2) \beta E (1/ \lambda_1
+ 1/ \lambda_2) \qquad {\rm for}
\quad i=1,2,\\[2mm]
{\rm H}_0 =- d^{2}/dr^{2} - (2/r) d/dr
- \lambda _1 (\lambda _1 - 1)/ r^{2} - 2 \beta
E /r +  (1/2) \beta E (1/ \lambda_1 + 1/ \lambda_2) ,\\[2mm]
\varepsilon = \beta E (1/ \lambda_1 + 1/ \lambda_2)
- 2(M^{2} - E ^{2}).
\ea
\end{equation}

Let us introduce  parasupercharges
${\rm Q}^{+}$ and ${\rm Q}^{-}$ of the form:
\[
{\rm Q}^{+} = \left( \begin{array}{ccc}0 & {\rm S}_1 & 0\\
0 & 0 & {\rm R}_2\\ 0 & 0 & 0 \end{array} \right), \qquad
{\rm Q}^{-} = \left( \begin{array}{ccc}0 & 0 & 0 \\{\rm R}_1 & 0 & 0\\
0 & {\rm S}_2 & 0 \end{array} \right),
\]
where
\begin{equation}
{\rm R}_i = d/dr + (1 + \lambda _i)/r - E \beta / \lambda _i ,
\qquad {\rm S}_i = - d/dr + (\lambda _i - 1)/r - E \beta / \lambda _i.
\end{equation}

{\samepage  Now it is straightforward to check that ${\rm Q}^{+}$,
${\rm Q}^{-}$, ${\rm Q}_1 = ({\rm Q}^{+} + {\rm Q}^{-})/2$, ${\rm
Q}_2 = (i/2)({\rm Q}^{+} - {\rm Q}^{-})$ and ${\rm H}$ satisfy the
commutation relations of the so-called  $p=2$
pa\-ra\-su\-per\-sym\-metric quantum mechanics of Rubakov and
Spiridonov (see e.g. [2, 4]):
\begin{equation}
\ba{l}
({\rm Q}^{\pm})^{3} = 0 ,\qquad {\rm Q}_i ^{3} = {\rm H }{\rm Q}_i,
\qquad
\lbrack {\rm Q}^{\pm},{\rm H} \rbrack = 0 ,\\[2mm]
\lbrace {\rm Q}_i ^{2}, {\rm Q}_{3-i} \rbrace +
{\rm Q}_i {\rm Q}_{3-i} {\rm Q}_i = {\rm H} {\rm Q}_{3-i},\qquad
i=1,2,
\ea
\end{equation}
where $[{\rm A, B} ] = {\rm A B} - {\rm B A}$, $\lbrace {\rm A,
B} \rbrace = {\rm A B} + {\rm B A}$.}

We see that the parasupercharges ${\rm Q}^{+}$, ${\rm Q}^{-}$ commute
with ${\rm H}$ and hence are (non-Lie) symmetries of our Coulomb
problem. Their existence is just the reason of the above-mentioned
threefold extra degeneracy.

\section{Conclusions and discussion}

Thus, in this paper we explained the threefold extra degeneracy of
discrete spectrum levels in a Coulomb problem for the modif\/ied
Stueckelberg equation for $k_2=0$, $g=2$, $j > 0$, $n > 1$ by the
presence of parasupersymmetry. Our results present a natural
generalization of supersymmetry in a Coulomb problem for the Dirac
equation [4].

It is also worth noticing that in the case of Coulomb f\/ield for
$k_2=0$, $g=2$ the equations of our model for the spin-1 states
with $j > 1$ coincide with the equations for the same states of
the Corben-Schwinger model [6] for $g=2$. Hence, their model also
is parasupersymmetric and possesses non-Lie symmetries --
parasupercharges ${\rm Q}^{+}$, ${\rm Q}^{-}$.

Finally, we would like to stress that, to the best of author's
knowledge, our model gives one of the f\/irst examples of
parasupersymmetry in a relativistic system with
non-oscillator-like interaction.

\medskip

{\it I am sincerely grateful to Prof. A.G.~Nikitin for statement of
the problem and sti\-mu\-la\-ting discussions.}

\label{sergeyev-lp}
\end{document}